\documentstyle[prd,aps,preprint,amsfonts,epsf]{revtex}
\draft
\tighten

\begin{document}

\preprint{
        \parbox{1.5in}{%
           hep-ph/yymmmmm \\
           CTEQ/PUB/02 \\
           FNAL/\ \ \ \ \ \  \\
           PSU/TH/136
        }
}

\title{Measuring Parton Densities in the Pomeron}

\author{John C. Collins$^{a}$, Joey Huston$^{b}$,
        Jon Pumplin$^{b}$, Harry Weerts$^{b}$,
        and J.J. Whitmore$^{a}$
}

\address{$^{a}$Department of Physics, Penn State University, University
            Park, PA 16802, U.S.A.
       \\
         $^{b}$Physics and Astronomy Department,
            Michigan State University, East Lansing MI 48824, U.S.A.
}

\date{December 8, 1994}

\maketitle

\begin{abstract}

   We present a program to measure the parton densities in the
   pomeron using diffractive deep
   inelastic scattering and diffractive photoproduction, and to
   test the resulting parton densities by applying them to other
   processes such as the
   diffractive production of jets in hadron-hadron collisions.
   Since QCD
   factorization has been predicted not to apply to hard diffractive
   scattering, this program of fitting and using
   parton densities might be expected to fail.  Its
   success {\em or failure} will
   provide useful information on the space-time
   structure of the pomeron.

\end{abstract}

%==================================================

\section {Introduction}
\label{sec:intro}

A spectacular effect in very high energy hadron collisions is
that of diffraction, wherein a hadron can scatter and emerge
unscathed, with only a small deflection and loss in longitudinal
momentum.  The rest of the final
state continues to exhibit the usual fragility of hadrons in a
collision where the total energy is far greater than all intrinsic
hadronic energy scales.  There is an old and established
quantitative phenomenology that describes diffraction and the
bulk of the cross section.  This is the theory of Reggeons and
the pomeron \cite{PomPhen}.
Unfortunately the understanding of the pomeron in
terms of the fundamental theory of QCD is weak.

With the advent of high energy colliders, and detectors with a
wide rapidity coverage, it is now possible to study diffractive
collisions that contain a hard scattering.  Such `hard diffractive
scattering' provides a tool whereby we
can
achieve a better
understanding of the pomeron at the fundamental level.

In the model of Ingelman and Schlein \cite{IS}, hard diffraction
is treated by considering the exchanged pomeron as an almost real
particle. This is entirely similar in spirit to the way low $Q^{2}$
electroproduction is used as a way of obtaining collisions with
almost real photons.
However, Collins, Frankfurt and Strikman \cite{CFS,FS,F} (CFS)
have questioned the validity of QCD factorization in this case,
and have given a specific mechanism for the
breakdown of factorization.  Their mechanism, `coherent hard
diffraction', has the signature that there will be events where the hard
scattering takes almost all the momentum of the pomeron.
Recent UA8 data \cite{UA8} appear to possess this property.
On the theoretical side, Levin and collaborators \cite{LevinPC}
have also verified that QCD predicts that factorization should
not apply to hard diffraction with two incoming hadrons, and a
calculation by Berera and Soper \cite{Berera} comes to the same
conclusion as CFS.

We therefore propose a program to measure the parton densities in
a pomeron.  Because the pomeron is an isosinglet and is
self-charge-conjugate, the number of independent distributions to
be measured is much smaller than for the proton.  We will show that
the program can therefore be carried out with existing detectors.
Two processes suffice to fix the parton densities, and a third
process will serve to test factorization.  See also a recent paper
by Ingelman \cite{Ingel93}.
Many of the individual points in this paper are already
known to experts in the field \cite{pompdf}.
Here we are showing the need for a coordinated program of measurements
involving different experiments.

If the UA8 results \cite{UA8} are indeed evidence of the CFS
mechanism \cite{CFS}, there will be a quite spectacular
failure of the program of fitting parton densities.  As CFS show,
this is a consequence of a failure of the applicability
of the factorization theorem, which is in turn a direct
consequence of the space-time structure of the pomeron.
Therefore, the success or failure of the fitting will
rather directly fulfill the promise of hard scattering to
turn a microscope onto the strong interaction.

Although the measurement of classical
diffraction requires ``Roman Pot''
detectors in the beam pipe, we will show that a substantial and
interesting part of our program can be
carried out without them.  One can instead use a rapidity gap
signature, thanks to the large rapidity coverage of current
detectors (CDF and D0 at the Fermilab $p \bar p$ collider,
H1 and ZEUS at the DESY $e^{-} p$ collider).  Indeed, ZEUS has
already reported data \cite{RapGapZEUS}
on events
with hard scattering in conjunction with rapidity gaps of this
kind.  These
data are quite suggestive of pomeron exchange.  In addition,
the D0 experiment has reported \cite{RapGapD0} on a search for jet
production with rapidity gaps {\it between} the jets; their data
are suggestive of the presence of such gaps.  CDF \cite{RapGapCDF}
has also found evidence for such gaps.

The outline of this paper is as follows.
In Sect.\ \ref{sec:processes} we explain the processes that are
of interest.  In Sect.\ \ref{sec:IS} we give the formulae
for the cross sections for these processes within the
Ingelman-Schlein model, and in Sect.\ \ref{sec:pdf.measure} we
show how to use them to measure the parton densities in the
pomeron.  In Sect.\ \ref{sec:rapgap} we show how
measurements can also be made with present detectors using
a rapidity gap signature.  Then in Sect.\ \ref{sec:nonfact} we
summarize the consequences of the CFS \cite{CFS} result for our
fitting program.  Some remarks on
experimental details are given in Sect.\ \ref{sec:checks}.
In Sect.\ \ref{sec:Processes}, we summarize the processes that
should be investigated.

%=====================================================

\section {The Processes}
\label{sec:processes}

We consider the following types of processes:
\begin{enumerate}

\item
Diffractive deep inelastic $e$-$p$ scattering: $e + p \to  e + X + p$.

\item
Diffractive deep inelastic $e$-$p$ scattering, with the measurement
of jets\footnote{
   By `jet', we will generally mean a jet that originates from the
   hard scattering, rather than a beam jet.
}
in the final state, {\it e.g.},
$e + p \to  e + J_{1} + J_{2} + X + p$.

\item
Diffractive photoproduction of jets, with the jets being produced by
the direct photon process: $\gamma + p \to J_{1} + J_{2} + X + p$.

\item
Diffractive photoproduction of jets via a resolved photon process,
{\it i.e.,} one with a photon remnant jet (a photon beam jet).

\item
Diffractive hadroproduction of one or more jets:
$p + \bar p \to  J_{1} + \dots + X + \bar p$.

\item
Other hard diffractive scattering in hadron-hadron collisions,
{\it e.g.}, production of the $W$ or the $Z$:
$p + \bar p \to  W/Z + X + \bar p$,
and production of direct photons at large transverse momentum:
$p + \bar p \to  \gamma  + X + \bar p$.

\end{enumerate}
In each of these processes we can allow for different numbers of
jets than those
explicitly indicated.  In each case the explicit final-state
$p$ (or $\bar p$) is a small angle diffracted proton (or anti-proton)
that carries $\gtrsim 95\%$ of the momentum of the incident $p$
(or $\bar p$).  The generalization in which the diffracted particle
is replaced by a low-mass inelastically diffracted state will be
discussed in Sect.\ \ref{sec:rapgap}.

For the sake of definiteness, the following discussion of the
kinematics will be presented for the case of hadron-hadron
scattering.  Only minor changes are required for the other
processes.

In terms of the momenta $p_{1}$ and $p_{2}$ of the incoming hadrons,
the diffracted hadron's momentum is
\begin{equation}
   {p_{2}'}^{\mu } = (1-x_{{\Bbb P}}) \, p_{2}^{\mu }
   + \epsilon  \, p_{1}^{\mu } + p_{T}^{\mu } \; ,
\end{equation}
with $x_{{\Bbb P}} \ll 1$ denoting the fraction of the diffracted
hadron's momentum transferred by the exchanged pomeron to the rest
of the final state, and $p_{T}^{\mu }$ denoting the
transverse momentum of
the diffracted hadron.
The quantity $\epsilon $ is very small
(${\cal O} (1/s)$),
and the square of the invariant momentum transfer is
\begin{equation}
 t = (p_{2}' - p_{2})^{2}
   = -(|p_{T}|^{2} + x_{{\Bbb P}}^{2} \, {m_{2}}^{2}) /
     (1-x_{{\Bbb P}}) \approx -|p_{T}|^{2} \;.
\end{equation}
For pomeron exchange to dominate, it is generally assumed that
one needs
\begin{equation}
   x_{{\Bbb P}} \lesssim 0.05 \;.
\end{equation}

Our definition of the pomeron is essentially kinematic:  it is
whatever exchanged object gives the dominant behavior at small
$x_{{\Bbb P}} = 1 - x$, where $x = {p_{2}'}^{+} / {p_{2}}^{+} \,$
is the fraction ($\approx 1$) of the large light-cone momentum
kept by the diffracted hadron.\footnote{
   We define light-cone coordinates for any vector $V^{\mu }$
   by $V^{\pm }=(V^{0} \pm V^{z})/\sqrt 2$ and $V^{T}= (V^{x},V^{y})$
   in a frame in which the
   incoming particle $p_{1}$ is moving in the $-z$ direction and
   $p_{2}$ is moving in the $+z$ direction.  See App.\
   \ref{sec:lc} for
   more details.  Our $z$ axis is the same as
   in ZEUS, when $p_{2}$ is the proton.
}
This is equivalent to defining
the pomeron as the mechanism that produces rapidity gaps, since
energy-momentum conservation implies a substantial rapidity
gap ($\gtrsim \ln(1/x_{{\Bbb P}})$) between the diffracted hadron
and the rest of the final state whenever $x_{{\Bbb P}}$ is small.

\begin{figure}
   \begin{center}
      \leavevmode
      \epsfxsize=0.3\hsize
      \epsfbox{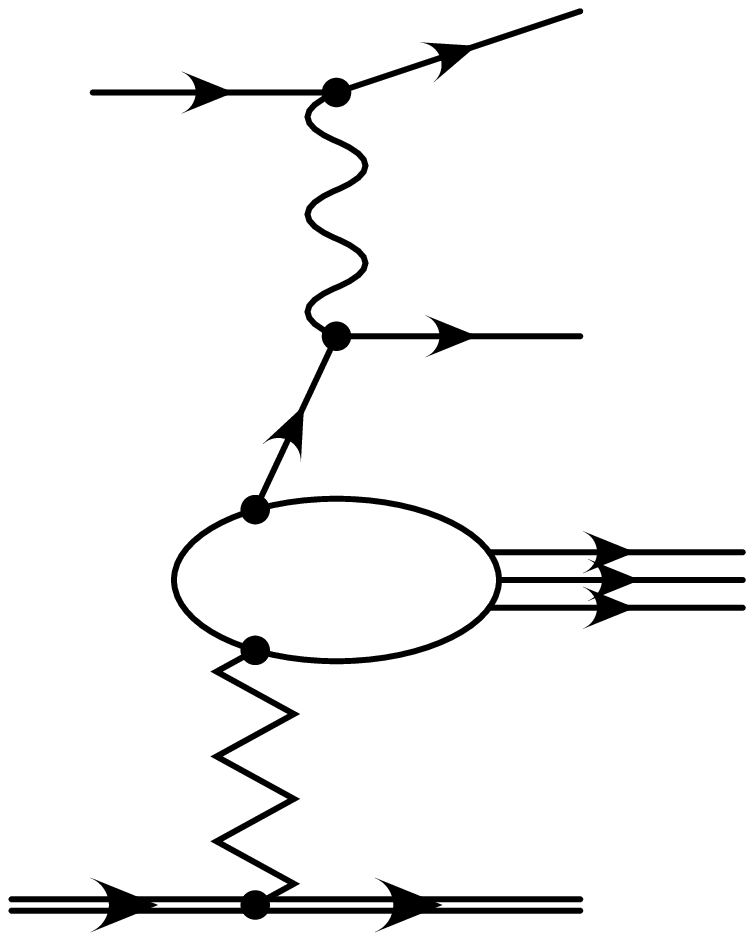}
      \hspace*{1cm}
      \epsfxsize=0.3\hsize
      \epsfbox{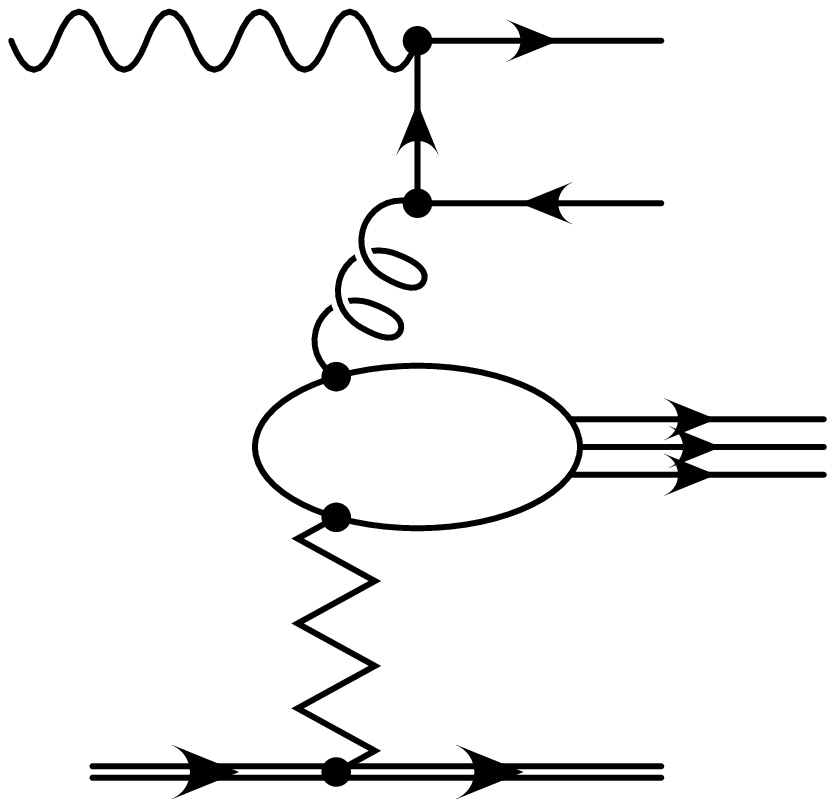}
   \\
      (a) \hspace*{6cm} (b)
   \\[1.5cm]
      \leavevmode
      \epsfxsize=0.3\hsize
      \epsfbox{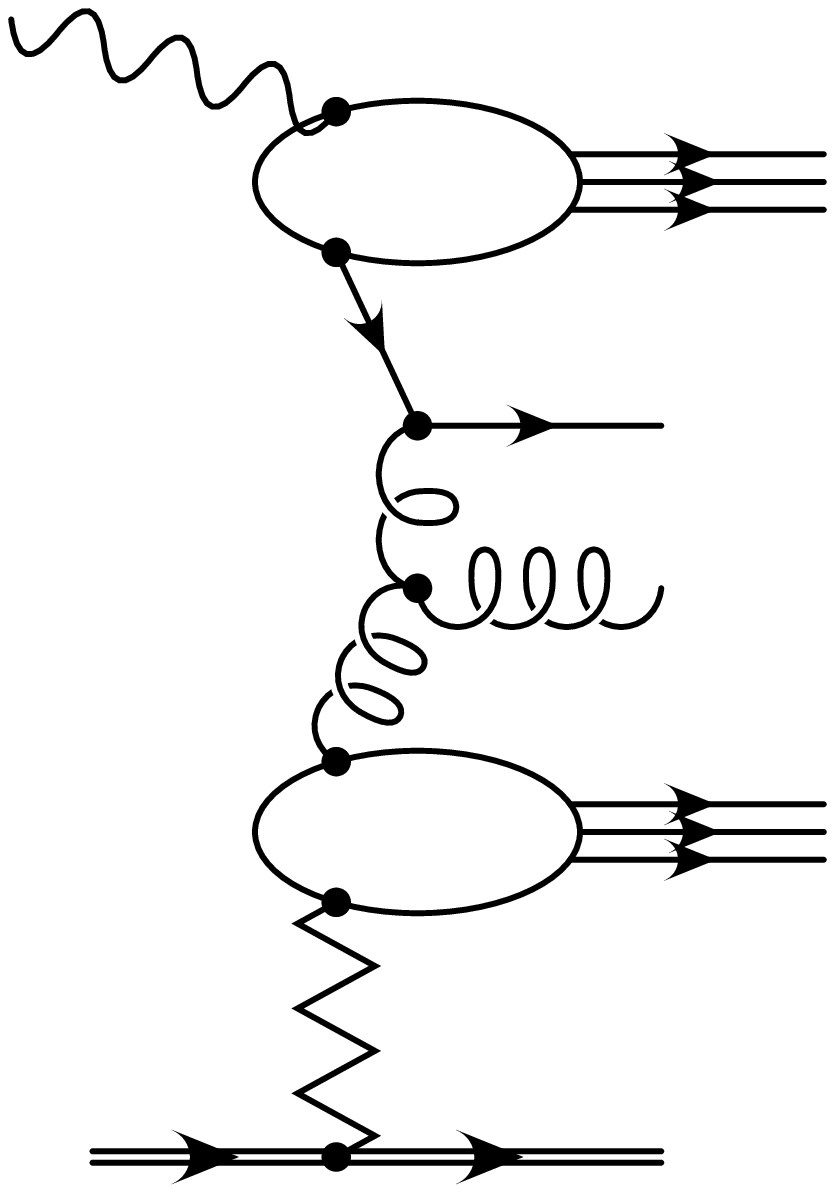}
      \hspace*{2cm}
      \epsfxsize=0.3\hsize
      \epsfbox{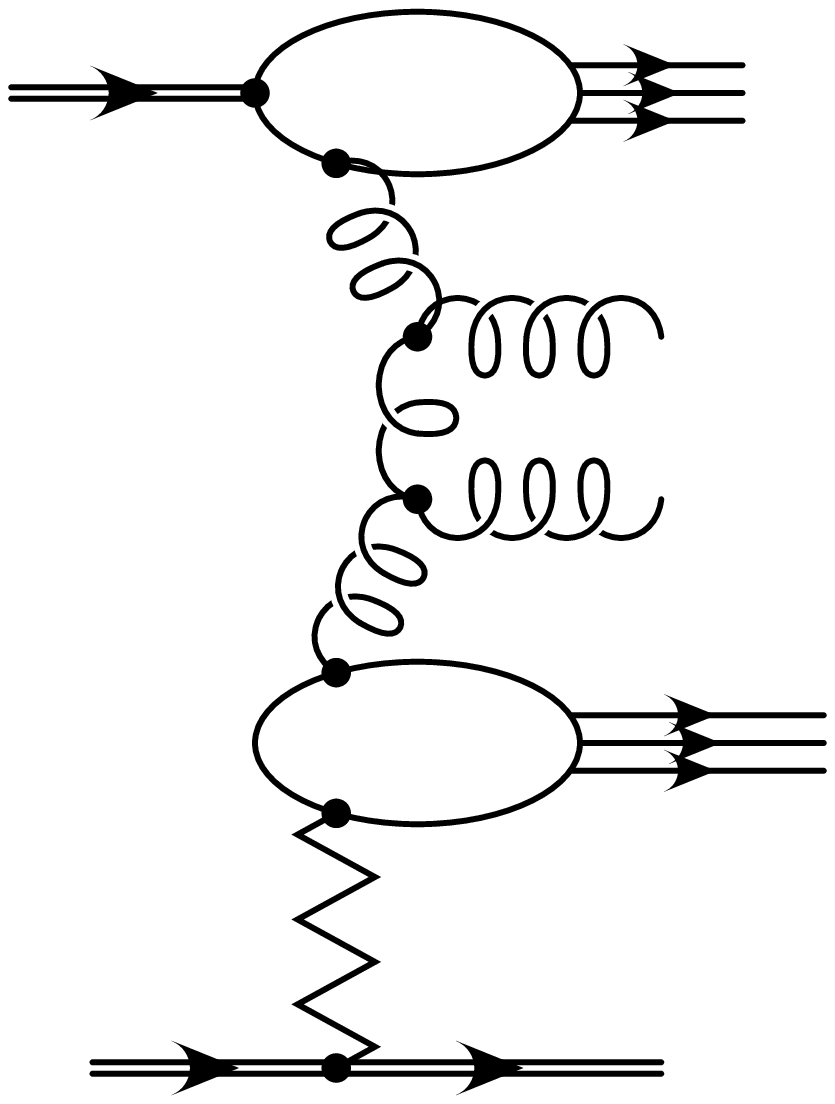}
   \\
      (c) \hspace*{6cm} (d)
   \end{center}
   \caption{The Ingelman-Schlein model for hard diffractive
   scattering, for the following processes: (a) electro-production,
   (b) direct and (c) resolved photo-production of jets, and
   (d) hadro-production of jets.
   }
   \label{fig:IS}
\end{figure}

Obvious graphs that should give leading twist contributions to
these processes are illustrated in Fig.\ \ref{fig:IS}.  In each case,
we show a typical lowest order graph for the hard scattering ({\it e.g.},
partons making jets), together with the pomeron exchange.  These
graphs represent the Ingelman-Schlein model, and the question we
wish to suggest be experimentally studied is whether the quantitative
phenomenology represented by that model is in fact correct.

%===================================================

\section {Ingelman-Schlein Model}
\label{sec:IS}

\subsection {Jets in diffractive hadron-hadron scattering}

Within the Ingelman-Schlein model, the cross section for
diffractive jet production in hadron-hadron collisions is
\begin{eqnarray}
   \frac {d\sigma (A+B\to J_{1}+\dots + B)}{dt \, dx_{{\Bbb P}}} &=&
   \frac {N}{16\pi } |\beta _{B{\Bbb P}}(t)|^{2} \,
       x_{{\Bbb P}}^{1-2\alpha _{{\Bbb P}}(t)} \,
       \sum _{a,b}\int dx_{a} \, \frac {dx_{b}}{x_{{\Bbb P}}} \,
\nonumber \\
  && \hspace{-1in}
     \times \,
     f_{a/A}(x_{a};\mu ) \, f_{b/{\Bbb P}}(x_{b}/x_{{\Bbb P}},t;\mu )
     \, \hat\sigma \left( a+b\to J_{1}+\dots  \right) \; .
\label{eq:DiffJet}
\end{eqnarray}
Here, $\mu $ is the usual renormalization and
factorization scale, which we will assume to be in the
$\overline{{\rm MS}}$
scheme, $f_{b/{\Bbb P}}(x_{b}/x_{{\Bbb P}})$ is the parton density
in the pomeron,
$f_{a/A}(x_{a})$ is the parton density in the non-diffracted hadron
$A$, and $\hat\sigma $ is the usual hard scattering cross section.
The last two quantities are identical to the corresponding quantities
used in hard non-diffractive scattering, so we consider them known.
The hard scattering cross section can be differential in the jet
variables, or integrated over a suitable range.   It can be
applied to any desired number of explicitly detected jets.

A tricky point is the normalization.  In view of the significance
of the test of the momentum sum rule for the parton densities in
the pomeron, this is important to get correct.  Unfortunately
there is controversy on the issue.  In Eq.~(\ref{eq:DiffJet}),
we have used\footnote{
   This represents a change from the preprint
   version of the paper.
}
the normalization factor from \cite{IS,BCSS}, but have introduced
a constant $N$ to allow for the possibility of a change.  In
any case, the idea is that the exchanged pomeron is to be thought
of as lying on a Regge trajectory, and thus is an analytic
continuation from particle poles in the region of $t\geq 0$.  The
existence of a momentum sum rule for a real particle is
unambiguous; the controversy is about the treatment of the
analytic continuation.

Ingelman and Schlein \cite{IS} would set
\begin{equation}
   N=1 \ \ \hbox{(Ingelman and Schlein)} .
\label{eq:ISNorm}
\end{equation}
Their argument relies on a formula given by Field and Fox
\cite{FieldFox} for the pomeron-proton total cross section.  This
formula, in effect, defines what a properly normalized pomeron
state would be.  The same formula for the pomeron-proton cross
section  is given, for example, by Kaidalov \cite{Kaidalov}, and
we believe it is to be regarded as the accepted formula in
Regge theory.

Landshoff \cite{DLnorm2} argues that a more
appropriate normalization would be one that works if the pomeron
exchange is replaced by a photon.  His argument leads to
\begin{equation}
   N=\frac {2}{\pi } \ \ \hbox{(Donnachie and Landshoff)}.
\label{eq:DLNorm}
\end{equation}

We have another argument for the incorrectness of
Eq.~(\ref{eq:ISNorm}).  We argue that we should also be able to
apply the Regge formula for diffraction when the exchanged
pomeron is replaced by a Regge trajectory on which there is a
massless spin-0 particle.  As we explain in App.\ \ref{sec:norm},
the normalizations are unambiguous and appear to give
disagreement with the standard Regge formulae
\cite{FieldFox,Kaidalov}.

At this point we find ourselves unable to resolve the
disagreements.  The cross section formulae in
\cite{FieldFox,Kaidalov} are given without proof, as if they are
obvious; but we find the proof far from obvious.  In any event,
the normalization factor $N$ is common to all the formulae for
diffractive scattering.

\begin{sloppypar}

We now return to the physics that is unaffected by the
normalization constant, which is everything but the momentum
sum rule.  The Ingelman-Schlein picture
(Eq.~(\ref{eq:DiffJet})) assumes dominance by single pomeron
exchange, which is a useful phenomenological approximation in
other applications of pomeron physics
\cite{PomPhen,DLnorm1,DLnorm2}.  It
also assumes the validity of factorization for hard scattering in
pomeron-hadron collisions --- the notion we wish to test.

\end{sloppypar}

The coupling $\beta _{B{\Bbb P}}(t)$ of the pomeron to the diffracted
hadron, and the pomeron trajectory $\alpha _{{\Bbb P}}(t)$, can be
obtained by fits to the elastic cross section at small $-t$:
\begin{equation}
   \frac {d\sigma (AB\to AB)}{dt} = \frac {1}{16\pi } \,
   |\beta _{A{\Bbb P}}(t)|^{2} \, |\beta _{B{\Bbb P}}(t)|^{2}
             \left(\frac {s}{s_{0}}\right)^{2\alpha _{{\Bbb P}}(t)-2}.
\label{eq:elastic}
\end{equation}
The pomeron coupling and its intercept also give total
hadron-hadron cross sections:
\begin{equation}
   \sigma _{{\rm tot}}(AB) = \beta _{A{\Bbb P}}(0) \beta _{B{\Bbb P}}(0)
         \left(\frac {s}{s_{0}}\right)^{\alpha _{{\Bbb P}}(0)-1} ,
\label{eq:sigtot}
\end{equation}
where we have omitted a signature factor that is reasonably close
to unity.

With $s_{0}=1\,{\rm GeV}^{2}$, we can use \cite{PomPhen,DLnorm1}
\begin{equation}
    \beta _{p{\Bbb P}}(t) = \beta _{\bar p{\Bbb P}}(t) \simeq
                4.6 \,{\rm mb}^{1/2}\,
                e^{ 1.9 \,{\rm GeV}^{-2} \, t } ,
\label{eq:proton-pomeron}
\end{equation}
and
\begin{equation}
    \alpha _{{\Bbb P}}(t) \simeq 1.08 + 0.25 \,{\rm GeV}^{-2} t \; .
\label{eq:alphaPom}
\end{equation}

Then Eq.~(\ref{eq:DiffJet})
can be used to measure the parton densities in
the pomeron --- or at least one combination of them.  The momentum
of the outgoing diffracted hadron
gives both $x_{{\Bbb P}}$ and $t$.  (A useful independent measure of
$x_{{\Bbb P}} = M^{2}/s$ can also be obtained from the diffractively
produced state, for events in which the edge of the rapidity gap
appears inside the detector --- see App.\ \ref{sec:lc}.)
Measurement of the
two jets allows one to reconstruct, to a first approximation, the
kinematics of the parton-parton collision.  In particular, it
gives the momentum fraction $x_{b}/x_{{\Bbb P}}$ of parton $b$ relative to the
pomeron.  QCD radiative corrections can be allowed for in the usual
fashion: one must only remember that the collision is to be treated
as a pomeron-hadron collision, whose energy will vary from event to
event.

\subsection{Diffractive deep inelastic scattering and photoproduction}

Diffractive hadroproduction of jets
by itself is not sufficient to extract the
flavor-separated parton densities.  We need to use more
processes.
It is very well established \cite{PomPhen} from conventional
Regge phenomenology that the pomeron is self-charge-conjugate and
isoscalar.
Thus the density of any flavor of anti-quark is equal to the
density of the corresponding quark,
\begin{equation}
   f_{i/{\Bbb P}}(x) = f_{\bar \imath/{\Bbb P}}(x),
\end{equation}
and the densities of the up and down quarks and antiquarks are
all equal:
\begin{equation}
   f_{u/{\Bbb P}}(x) = f_{d/{\Bbb P}}(x) = f_{\bar u/{\Bbb P}}(x)
   = f_{\bar d/{\Bbb P}}(x) \equiv  f_{q/{\Bbb P}}(x).
\end{equation}

To a good first approximation, there are therefore only two parton
densities to measure: the gluon density $f_{g/{\Bbb P}}$ and the
quark density $f_{q/{\Bbb P}}$.  The strange quark density
$f_{s/{\Bbb P}}=f_{\bar s/{\Bbb P}}$ is probably less important,
but it could be measured using charged-current charm
production at HERA, just as the strange quark density in nucleons
is measured by dimuon production in neutrino scattering.  The
other quarks (charm and bottom) are generally considered heavy
enough that their densities can be correctly generated
dynamically by evolution from the known light parton densities,
at least to an accuracy that is sufficient for our purposes.

Formulae similar to, but simpler than, Eq.~(\ref{eq:DiffJet})
can be written for
the diffractive deep inelastic structure functions and for the direct
photoproduction of jets.  If we measure the deep inelastic structure
functions $F_{1}$ and $F_{2}$ differentially in the momentum
of the diffracted hadron, they can be expressed in terms of
the structure
functions $F_{1}^{{\Bbb P}}$ and $F_{2}^{{\Bbb P}}$  for the
pomeron:
\begin{eqnarray}
\frac {dF_{1}^{\rm diffractive}(x_{\it bj},Q; t,x_{\Bbb P})}{dt\,dx_{\Bbb P}}
    &=& \frac {N}{16\pi } \, |\beta _{B{\Bbb P}}(t)|^{2} \,
    {x_{{\Bbb P}}}^{-2\alpha _{{\Bbb P}}(t)} \,
    F_{1}^{{\Bbb P}}(x_{\it bj}/x_{{\Bbb P}},Q)
\label{eq:DiffF1}
\\
\frac {dF_{2}^{\rm diffractive}(x_{\it bj},Q;t,x_{\Bbb P})}{dt\,dx_{\Bbb P}}
    &=& \frac {N}{16\pi } \, |\beta _{B{\Bbb P}}(t)|^{2} \,
    {x_{{\Bbb P}}}^{1-2\alpha _{{\Bbb P}}(t)} \,
    F_{2}^{{\Bbb P}}(x_{\it bj}/x_{{\Bbb P}},Q) \;.
\label{eq:DiffF2}
\end{eqnarray}
Note the different powers of $x_{{\Bbb P}}$ in the above
equations!  The pomeron coupling and trajectory function are
given by the expressions
in Eqs.~(\ref{eq:proton-pomeron}) and
(\ref{eq:alphaPom}).
Ordinarily, the structure functions $F_{1}$ and $F_{2}$ are defined for
totally inclusive deep inelastic scattering.  But our diffractive
cross section is actually semi-inclusive, so that there are extra
structure functions.  We explain these issues in App.\
\ref{sec:DiffStr}.

Similarly, for the direct photoproduction of jets we have
\begin{eqnarray}
     \left.
       \frac {d\sigma (\gamma +B\to J_{1}+\dots + p)}{ dt \, dx_{{\Bbb P}}}
     \right|_{{\rm direct}}
    &=& \frac {N}{16\pi } \, |\beta _{B{\Bbb P}}(t)|^{2} \,
    {x_{{\Bbb P}}}^{1-2\alpha _{{\Bbb P}}(t)}
\nonumber \\
    && \hspace{-1.5in}
       \times  \sum _{b} \int  \frac {dx_{b}}{x_{{\Bbb P}}} \,
       f_{b/{\Bbb P}}(x_{b}/x_{{\Bbb P}},t;\mu )
       \, \hat\sigma \left( \gamma +b\to J_{1}+\dots
            \right)_{{\rm direct}} .
\label{eq:DirJet}
\end{eqnarray}

The formula for the resolved contribution to the photoproduction
of jets is identical to that for hadroproduction,
Eq.~(\ref{eq:DiffJet}),
with the parton densities in the hadron $A$ replaced by the
parton densities in the photon.

It is also possible to consider the case where the {\em photon} is
diffracted quasi-elastically into a $\rho ^{0}$ meson, or to some other
low mass state that carries $\gtrsim 90\%$ of the original
$\gamma $ momentum, with the corresponding pomeron undergoing
hard scattering on the other incident particle.  See
\onlinecite{Brodskyetal} for some recent predictions.

%==========================================

\section {Measuring the Parton Densities}
\label{sec:pdf.measure}

For the sake of definiteness,
we consider diffractive deep inelastic scattering
(DIS) and diffractive direct
photoproduction of jets as the processes to measure the parton
densities of the pomeron.  We then treat the other
processes of Sect.\ \ref{sec:processes}
as providing a tests of factorization.  This is reasonable, since
the arguments against factorization apply at their strongest to
hard diffraction in hadron-hadron collisions, and thus to
production of jets both in diffractive hadron-hadron collisions
and in diffractive resolved photoproduction.

With flavor-separated parton densities for the pomeron, one can
test the momentum sum-rule:
\begin{equation}
    \sum _{i} \int _{0}^{1} dx \, x \, f_{i/H}(x) = 1,
\label{eq:MomSR}
\end{equation}
as applied to the pomeron.  In the case of parton densities
in an ordinary hadron, this sum rule is a prediction (provided
that a suitable definition of the parton densities is used, such
as the $\overline{{\rm MS}}$ definition).  But it is
by no means required that the sum rule hold for the parton
densities in the pomeron \cite{BCSS,DLnorm2,LevinPC}.  The reason is that
the proof of the momentum sum rule Eq.~(\ref{eq:MomSR})
requires one to
take the expectation value of the energy-momentum tensor in the
hadron state $H$.  For the proof to apply to the pomeron, one
would have to construct a quantum mechanical state for the
exchanged pomeron in diffractive processes, and such a concept
does not appear to exist.

Deep inelastic electron-proton scattering primarily measures the
quark distribution:
\begin{eqnarray}
    F_{2}^{{\Bbb P}}(x) &=& x \sum  e_{i}^{2} f_{i/{\Bbb P}}(x)
    + {\cal O}(\alpha _{s})
\nonumber\\
              &=& \frac {10}{9} xf_{q/{\Bbb P}}(x)
    + \frac {2}{9}xf_{s/{\Bbb P}}(x) +  {\cal O}(\alpha _{s}) \;.
\label{eq:DIS0}
\end{eqnarray}
The strange quark term in Eq.~(\ref{eq:DIS0})
is presumably substantially smaller than the up and
down quark terms, and can be ignored in a first pass at
a fit.  However, the gluonic part of the ${\cal O}(\alpha _{s})$
term is probably significant at small $x_{\it bj}/x_{{\Bbb P}}$.

To measure the strange quark distribution, one could investigate
the charged current process with the production either of muons or
of charm in the current fragmentation region:
\begin{equation}
   e + p \to  \nu  + \mu /D + X + p.
\end{equation}
Muons would typically have come from charm decay, and charm
production is mostly from scattering off strange quarks, in
charged current events.

Direct photoproduction of jets would be sensitive to both the
quark and gluon densities.
The version of Eq.~(\ref{eq:DirJet})
that applies to this case is, schematically,
\begin{equation}
   \sigma (\gamma +p\to \hbox{jets} + X + p) \Big|_{\rm direct}
   \propto  \alpha _{{\rm em}} \, \alpha _{s}
  \left( A \, \frac {10}{9} \, x \, f_{q/{\Bbb P}} +
            A \, \frac {2}{9} \, x \, f_{s/{\Bbb P}}
                        + B \, x \, f_{g/{\Bbb P}}
                        + {\cal O}(\alpha _{s})
                 \right),
\label{eq:Dir.Jet.Schematic}
\end{equation}
where $A$ and $B$ are known coefficients that carry the detailed
kinematic dependence from the pomeron factors and the hard
scattering coefficient $\hat \sigma $ in Eq.~(\ref{eq:DirJet}).  The
formulae for these coefficients are the same as for the same
processes without the diffractive requirement.
The photon does not have to be real:
identical physics would be probed by DIS
with an extra jet
beyond the minimum current quark jet given by the parton model.
The only penalty would be a loss
of cross section.  The simplest way to extract the parton
densities in Eq.~(\ref{eq:Dir.Jet.Schematic}) would be to measure
2-jet cross sections, with the kinematics of the
underlying photon-parton scattering being deduced by treating it
as a $2 \to 2$ scattering.

Hence the diffractive subsets of the two processes,
DIS and direct photoproduction of jets,
will measure the parton densities.  The results of a fit can then
be put into Eq.~(\ref{eq:DiffJet})
to predict diffractive cross sections
for other hard processes, in particular
hadroproduction of jets and of $W$ or $Z$ bosons
at the Tevatron, and diffractive resolved
photoproduction of jets at HERA.  They can also be used to
calculate the cross section that UA8 \cite{UA8} have measured.  If
the UA8 results are interpreted according to the CFS
mechanism \cite{CFS}, then
a readily visible excess of events can be predicted
at $x_{b}/x_{{\Bbb P}}\to 1$ for the hadron-hadron experiments relative to the
DIS and direct photoproduction experiments.  In any event the parton
densities measured at HERA can be compared with those fitted
by UA8 to their data, both in shape and in absolute
normalization.

A full quantitative analysis and fit for the parton densities in
the pomeron would require the use of the higher order corrections
to the formulae for the cross sections and structure functions,
such as the ${\cal O}(\alpha _{s})$ terms in Eq.~(\ref{eq:DIS0})
and Eq.~(\ref{eq:Dir.Jet.Schematic}).  These are standard known
expressions.

%==========================================

\section {Rapidity Gap Signature}
\label{sec:rapgap}

The ideal experimental signature of the pomeron for the purposes of
deep inelastic physics in a colliding beam experiment
involves observing a ``quasi-elastically''
scattered proton using a Roman Pot detector.  These detectors
should be
located very close to the unscattered beam direction,
because, in the region of pomeron-dominance,
the quasi-elastic protons have longitudinal momenta
$0.9$ -- $0.99$ or more
times the incident proton momentum, and transverse
momenta $\lesssim 1 \, {\rm GeV} \,$.  It then follows by kinematics
that there is a rapidity gap between the quasi-elastic proton and
the rest of the event.%
\footnote{
   When the pomeron's momentum fraction $x_{{\Bbb P}}$ is very small, the
   resolution of the measurement of the proton's momentum may not
   be sufficient to get a good determination of $x_{{\Bbb P}}$.  In that
   case, measurement of the other detected particles will serve to
   determine $x_{{\Bbb P}}$ by the relation
   $x_{{\Bbb P}}=M_{X}^{2}/s$, a measurement which
   will be dominated by the particles close to the edge of the
   rapidity gap, and by particles with high transverse
   momentum --- including the scattered electron in
   deep-inelastic events.  The use of light-cone coordinates
   will facilitate this computation --- see App.\ \ref{sec:lc}.
   It should be emphasized that the scattered electron at HERA is
   often in a kinematic configuration that is completely
   analogous to the isolated proton in diffractive scattering,
   and that measurement of the rest of the final state has proved
   to be in practice an effective method to determine the
   quantity corresponding to $x_{{\Bbb P}}$.
}

Roman Pots have been added to the HERA detector ZEUS, but are not
yet present in H1.   They are also not present in either detector at the
Tevatron, although CDF took some early data with pots, and has proposed
adding new ones in an eventual upgrade.  In all current
experiments, there are several units of rapidity between the
hadrons that can be detected by a Roman Pot detector and by the rest
of the detector.

Fortunately, much of the same physics can be probed using a rapidity
gap signature alone.  The gap signature is defined by a requirement
that no final state particles are produced in a
sufficiently long region
in pseudorapidity that
extends from the edge of
the main detector toward its center.
The data from \cite{RapGapZEUS,RapGapH1}, together with standard
pomeron phenomenology, suggest that ``sufficiently long'' can be
read as ``length $\gtrsim 3$ units'' of (pseudo-)rapidity.
The gap
signature (for a proton beam) allows a mixture  of
(1) single ``quasi-elastic'' protons,
(2) low-mass diffractive excitations of the proton, and
(3) moderately high mass diffractive excitations, whose mass is
nevertheless low enough that the entire state misses the detector,
to disappear in or close to the beam pipe.  Diffractive events
have already been observed using this technique at ZEUS
\cite{RapGapZEUS} and H1 \cite{RapGapH1}.

\begin{figure}
   \begin{center}
      \leavevmode
      \epsfxsize=0.3\hsize
      \epsfbox{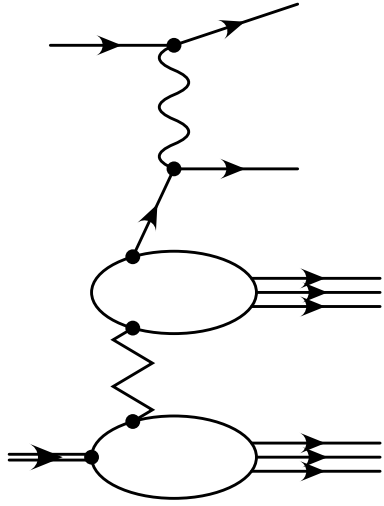}
   \end{center}
   \caption{Regge diagram for hard diffractive DIS with
      low mass excitation of proton.}
   \label{fig:RapGap}
\end{figure}

Cases (2) and (3) correspond to replacing the lower pomeron-proton
vertex in Fig.\ \ref{fig:IS} by a diffractive excitation as in
Fig.~\ref{fig:RapGap}.  The factor
$|\beta _{B{\Bbb P}}(t)|^{2}$ in Eq.~(\ref{eq:DiffJet}) is replaced
by a factor from ordinary soft diffraction:
\begin{equation}
    |\beta _{B{\Bbb P}}(t)|^{2} \longrightarrow
    |\beta _{B{\Bbb P}}(t)|^{2} \left( 1 + \Delta _{B}(t) \right) ,
\label{eq:correction.factor}
\end{equation}
where the correction factor $1+\Delta $ can obtained from the ratio of
diffractive to elastic scattering:
\begin{equation}
   \frac {d\sigma ^{{\rm diff}}(A+B\to A+X)}{dt} =
   \frac {d\sigma ^{\rm el}(A+B\to A+B)}{dt}
   \left( 1 + \Delta _{B}(t) \right) .
\label{eq:diff.el}
\end{equation}
The cross section we use here for single diffraction has an
integral over the mass of the diffractively excited system (the
lower group of final-state hadrons in Fig.~\ref{fig:RapGap}).
Thus the correction factor $1+\Delta $ will depend on the limit that is
placed on this system.  In the case of the rapidity gap
signature, this limit is given by requiring that the hadrons
be within the beam-hole of the detector.

The use of (\ref{eq:correction.factor}) assumes the validity of Regge
factorization in its simplest form for ordinary diffraction,
elastic scattering and for hard diffractive scattering.  Thus
it predicts that the ratio of contributions from
quasi-elastic scattering and diffractive excitation in rapidity
gap events should be the same in DIS events as it is in low $p_{T}$
scattering with pure quasi-elastic scattering of the oppositely-directed
beam particle.   We can therefore use the CDF Roman Pot data on
single diffraction \cite{CDFdiff} to estimate what goes down the beam pipe
in events with a gap trigger.   Using CDF's parameterization of their
data, we estimate the cross section for
$\bar p + p \rightarrow \bar p + p^{*}$ in which the $p^{*}$ has such a
low invariant mass ($\lesssim 6 \, {\rm GeV}$)
that it
would completely miss a detector confined to $| \eta | < 4$, to be
approximately $3 \, {\rm mb}$.
Meanwhile the integrated elastic cross section
$\bar p + p \rightarrow \bar p + p$ is
approximately $20 \, {\rm mb}$,
so a rapidity gap trigger would correspond $87\%$ of the
time to a single $p$ going down the beam pipe.

The data on inelastic diffraction are actually somewhat inconsistent
with regard to normalization \cite{O}.  This partly reflects
differences in the kinematic ranges included by different experiments.
More detailed analysis, with more data from modern colliders, is needed.
Analysis \cite{newdiff} of diffractive data also does not give a
pomeron trajectory identical to that obtained from data on
total and elastic cross sections, so there may be significant
contributions from multi-pomeron exchange.

It will be important to compare the rapidity gap trigger
with the pure quasi-elastic diffractive
trigger, using the ZEUS Roman Pots.  This will
provide a ``renormalization factor'' that can be used to relate
the two types of trigger in other experiments where
only the rapidity gap method is available.  The observed value of this
factor, and the possibility that it may vary with the deep inelastic
kinematic variables, or be different for a low $p_{T}$ process such as
quasi-elastic $\rho ^{0}$ or $J / \psi $ production,  will provide an
interesting test of Regge factorization for the pomeron.

Meanwhile, if Regge factorization is at least close to the truth, the
``renormalization factor'' will be somewhat less than unity,
corresponding to
small contributions from the inelastic $p^{*}$ states, so it
should be possible to estimate the quasi-elastic diffractive cross
sections rather accurately from data based on the rapidity-gap method.

Whereas the measurement of a single diffracted hadron in the
beam pipe allows one to measure a diffractive cross section that is
differential in $t$, the rapidity gap signature allows only an
integral over $t$ to be measured.  This is sufficient for a measurement
of parton densities in the pomeron, although it does not give information
on the interesting subject of the $t$-dependence.
It is convenient in this regard that the relevant detectors
(CDF, D0, H1, and ZEUS) have very similar kinematic
coverage relative to the proton beam pipe:  Parton densities measured
by the gap method, which implicitly contain a weighted average over $t$,
will be directly comparable, with small corrections.

Significant Regge
phenomenology would be required, however, to relate these measurements
to those of UA8, who have a true quasi-elastic requirement with $-t$
in the range of $1$ to $2 \, {\rm GeV}^{2} \,$.  Normal soft-pomeron
phenomenology cannot necessarily be applied at such large $|t|$.
For example, we observe that there is a substantial change in the
behavior of the $p \bar p$ elastic cross section below and above
$-t\approx 1\,{\rm GeV}^{2}$ \cite{elastic}.  Ordinary pomeron phenomenology
works well only below this boundary.  Above it, a Regge analysis
is still appropriate ($p_{T} \ll \sqrt s$); but possibly the
pomeron becomes more like the perturbative pomeron of BFKL \cite{BFKL},
than the conventional phenomenological soft pomeron.

There is no fundamental requirement that the parton densities in
the pomeron must be independent of $t$.  But the strongest $t$ dependence
is presumably in $\beta _{B{\Bbb P}}(t)$, since that represents an
elastic coupling. Any experimental information on the
$t$-dependence would be welcome.\footnote{Compare
   in particular the recent work by Sotiropoulos and
   Sterman \cite{SotSt} concerning the $t$-dependence of elastic
   scattering.  }
It is also possible that the amount of the breakdown of
factorization, if it occurs, is $t$-dependent.

%==========================================

\section {Non-factorization}
\label{sec:nonfact}

\subsection{Delta-function terms}

CFS \cite{CFS} predict that
there are non-factorizing leading twist terms in hard diffractive
scattering, but only in certain processes.  These terms behave
kinematically as if the pomeron had a point-like
component, like a photon.  For example there should be such a
term in diffractive jet production in hadron-hadron collisions,
where it will appear as a peak in the
cross section at $x_{b}/x_{{\Bbb P}}\approx 1$.  In the simplest
approximation to the theory the term is a delta-function,
proportional to $\delta (x_{b}/x_{{\Bbb P}}-1)$.  However, the
strength of the delta-function is not universal between different
processes.

A simple way to look for the CFS mechanism is to reconstruct
the values of the parton momentum fractions
$x_{a}$ and $x_{b}/x_{{\Bbb P}}$
from the jet momenta in each event.  On an event-by-event basis,
the kinematic analysis should be done relative to the
pomeron-proton center-of-mass frame, or by the use of light-cone
coordinates.  Then one would plot the
cross section as a function of $x_{b}/x_{{\Bbb P}}$.  If the CFS
mechanism is valid, there should be a contribution peaked at
$x_{b}/x_{{\Bbb P}}\approx 1$.

A full analysis must recognize that the jets will be smeared out,
compared to the situation at the parton level.  Thus, as in the
UA8 analysis \cite{UA8},
either substantial unfolding is necessary, or the data and theory
should only be compared after smearing the theory.
Smearing by hadronization and detector resolution is much
more noticeable for a delta-function than for a smooth function!
Moreover, there will be a correction due to higher-order
perturbative QCD corrections to the jet production, including
production of 3 or more jets.  A full analysis could best be done
by proceeding from assumed parton densities, and including
resolution, hadronization, higher order QCD, etc., to calculate
a cross section to be compared with the data.  One would then
adjust the parton densities to fit the data.  This analysis will
of course have limited power to distinguish a genuine
delta-function from a narrow peak, since the physical cross section
will not display a true delta-function.\footnote{
   In fact, when perturbative QCD predicts a cross-section to
   have a delta-function, or some other singular behavior, the
   prediction should be interpreted in the sense of the theory of
   distributions --- see Sect.\ \ref{subsec:deltafn}.
}

\subsection{The $\delta $ function is not exact}
\label{subsec:deltafn}

It is only in the simplest approximation that the CFS coherent
term is proportional to a delta-function at $x_{b}=x_{{\Bbb P}}$.  Higher
order corrections are likely to smear it into something that is
merely strongly peaked at $x_{b}\approx x_{{\Bbb P}}$.  Experimental
resolution and
jet hadronization will smear it even more.  The real experimental
signatures are that diffractive jet production should be much
larger as $x_{b}\to x_{{\Bbb P}}$ than can be explained by
anything but a density
of partons in the pomeron that rises strongly in that region.
CFS\cite{CFS} predict
that any similar excess in DIS is higher twist, i.e., that in DIS
the fraction of the coherent (or super-hard) events
falls off as a power of $Q$ as $Q$ increases.

The meaning of a delta-function in any calculation of a hard
process is not that a physical cross section has the
delta-function, but that the experimental and calculated cross
sections should agree, up to higher-twist terms, only after they
are both averaged with a smooth test function.  That is, the
theoretical prediction should be interpreted in the sense of
distributions.  (Compare Ref.\ \onlinecite{Tkachov}.)
In our case that
means that the actual prediction is simply that there is a
substantial excess in a narrow range of $x_{b}/x_{{\Bbb P}}$
close to $1\,$.

\subsection{Other non-factorization}

The CFS mechanism is a particularly clear example of the failure
of the factorization theorem.  In general, to prove
\cite{DEL,CW,fact} the factorization theorem for a
large-momentum-transfer process in hadron-hadron scattering, one
must require that the cross section be inclusive.  This
requirement is violated by a diffractive condition on the final
state.  Besides the CFS term, there can be a general failure of
the factorization theorem which would manifest itself in a
failure of the universality of the parton densities: a
simultaneous fit of parton densities to many processes would
fail.

The requirements of the final-state cancellation are much less
stringent in deep inelastic scattering---see \onlinecite{CFS,fact}.
So it is quite possible
that the factorization theorem holds within that class of
processes---more theoretical investigations are needed on this
point.  This restricted factorization theorem
would be manifested by a universality of the
parton densities within this subset of processes.
Such processes include
the diffractive components of $F_{1}$ and $F_{2}$ at different values
of $Q$, and the production of various numbers of jets in DIS.  (In
jet production in DIS, one should require the jet(s) to have
transverse momentum of order $Q$, rather than much larger than
$Q$.  Otherwise one can get
something more like photoproduction, where factorization should
fail, since photoproduction is just a special case of
hadron-hadron scattering for the purposes of the factorization
theorem.)

\subsection{Process dependence}

The coherent term predicted by CFS is
a contribution not present in the Ingelman-Schlein model, and
is predicted to be quite different in different processes.  Thus
data for a suitable set of processes should not be consistently
fit by a single set of parton densities, at least for the
delta-function term.  It is also not necessarily true that the
continuum term in the parton densities is universal, or that the
$t$-dependence of the delta-function term is similar to the
continuum (Ingelman-Schlein) term.

\begin{figure}
   \begin{center}
      \leavevmode
      \epsfxsize=0.3\hsize
      \epsfbox{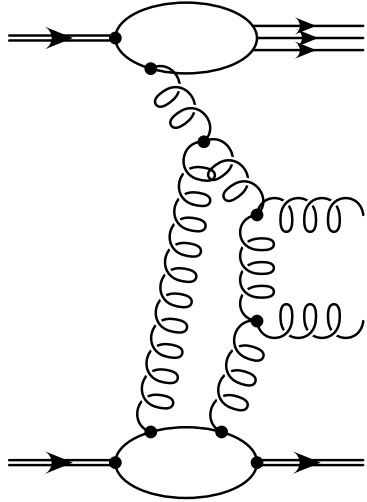}
   \end{center}
   \caption{Example of low-order graph for model of
      Collins, Frankfurt and Strikman \protect\cite{CFS}.}
   \label{fig:Coh}
\end{figure}

To understand the process-dependence, one must understand CFS's
reasoning.  They model the pomeron by two-gluon exchange as in
Fig.~\ref{fig:Coh}.  One of the gluons (the `hard
gluon') carries most of the longitudinal momentum of the pomeron
and the other gluon is relatively soft, at least in terms of its
longitudinal momentum. The soft gluon is to be regarded as
exchanged between two oppositely moving gluons (or a gluon and a
quark), and one of these gluons is the `hard gluon' in the
pomeron.  Since a diffractive requirement has been imposed on the
final state, the usual cancellation \cite{DEL,CW,fact}
of the effects of soft gluons, which is responsible for QCD
factorization in inclusive processes, no longer occurs.

\begin{figure}
   \begin{center}
      \leavevmode
      \epsfxsize=0.3\hsize
      \epsfbox{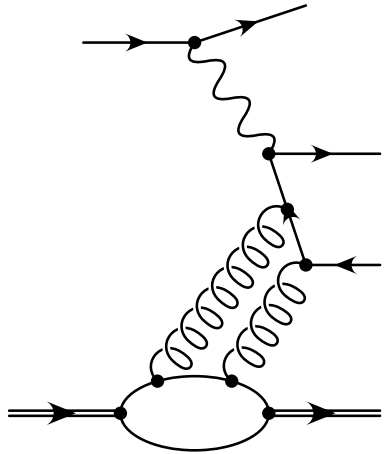}
   \end{center}
   \caption{Graph for coherent pomeron in deep inelastic
      scattering.}
   \label{fig:CohDIS}
\end{figure}

Now a gluon only couples to color fields, so the corresponding
graphs do not exist in DIS, where an incoming electron plays the
role of the uppermost gluon in Fig.~\ref{fig:Coh}.
Instead, in DIS, one must consider graphs like
Fig.~\ref{fig:CohDIS}, as has
been done by Donnachie and Landshoff \cite{DLHERA} and by Nikolaev
and Zakharov \cite{NZ}.
In order for the kinematics to correspond to the
coherent pomeron term,
we assume that the quark and anti-quark coming
from the virtual photon have transverse momenta comparable to the
virtuality $Q$ of the photon.  The quark and anti-quark thus give
rise to
jets to be associated with the hard scattering.  Graphs like
Fig.~\ref{fig:CohDIS} exist and give an approximate
delta-function, just like the graphs in diffractive
hadroproduction. But since the
second gluon of the pomeron attaches directly to the hard
scattering, or to an out-going parton, the result is
higher twist \cite{CFS}.  That is, if one analyzes the cross section in terms
of a continuum plus a delta-function term, then the delta-function
falls off as a power of $Q$ relative to the continuum.
This is in contrast to the delta-function term in the
hadroproduction of jets, which is predicted by CFS to scale.
CFS note that the cross section calculated by Donnachie and
Landshoff \cite{DLHERA} agrees with this observation.

\subsection{Systematics}

We summarize here the systematics of the predicted delta-function
terms:
\begin{enumerate}

\item In DIS and the direct photoproduction of jets, the
    delta-function term should be higher twist, {\it i.e.},
    it should die away as a power of $Q$ or of jet transverse
    energy $E_{T}$ when one tries to make a scaling test.

\item In hadroproduction of jets, the delta-function should be
    of leading twist: it will scale and should not
    fall off as a power of increasing jet $E_{T}$.

\item In $W$ and $Z$ production, there should be at most a small
   delta-function term when $q_{T}$, the transverse momentum of
   the $W/Z$, is small.  But as one increases $q_{T}$, a jet will
   appear.
   If one analyzes the process in terms of the kinematics of the
   $W/Z$ plus jet(s) system, there should
   be a delta-function term which will scale.\footnote{
       The reason for needing an extra jet in $W/Z$ production to
       get the delta-function term is that the hard process
       without the jet is ${\Bbb P} +q \to  W/Z$.  The only way to
       balance the baryon number is to emit a soft quark into the
       final state, which will be suppressed.
   }

\item {\em Resolved} jet production in diffractive photoproduction
   should also have a leading-twist
   delta-function, at least if there is sufficient
   kinematic range to allow the process.

\end{enumerate}
How to determine experimentally whether some diffractive quantity
is leading or higher twist is explained in
Sect.~\ref{sec:checks}.

In addition, there may be substantial $t$-dependence.  The most
reliable prediction of the coherent delta-function term is at
sufficiently large $|t|$ that the perturbative model is at
least qualitatively valid.  At low $|t|$ --- an unambiguously
non-perturbative region for the pomeron --- one cannot trust the
two-gluon exchange model.

When a rapidity gap signature is used, there is an integral over
the unmeasured value of the pomeron's momentum transfer.
This integral is dominated by very low
$|t|$ --- certainly well below $1 \,{\rm GeV}^{2}$.
Hence one cannot necessarily expect perfect
consistency between the jet
production reported by UA8 \cite{UA8} and the jet production that
would be measured by CDF and D0 using a rapidity gap signature
for pomeron exchange.  {\em The difference itself is an important
probe of the space-time structure of the pomeron.}  Of course, a
test of the CFS mechanism is that there should be consistency
between the excess of events at large $x_{b}/x_{{\Bbb P}}$ measured
by UA8, and
a similar excess that could be measured by CDF if
Roman Pot detectors are reinstalled and used in a kinematic region
corresponding to that of UA8.

%==========================================

\section{Consistency Checks: Scaling Tests}
\label{sec:checks}

\subsection{Pomeron exchange}
\label{subsec:check.pomeron}

To check that a measurement of a hard diffractive process is really
associated with pomeron exchange, one must verify the $x_{{\Bbb P}}$
dependence.  We are in effect
defining the pomeron as whatever is responsible
for the leading power behavior as $x_{{\Bbb P}}\to 0$.  For the name `pomeron'
to be appropriate, the intercept, $\alpha _{{\Bbb P}}(0)$,
must be close to (or above) unity,
the approximate value that appears in soft processes like the
total cross section for hadron-hadron scattering.

Diffractive cross sections go like
$d\sigma /dx_{{\Bbb P}} \propto x_{{\Bbb P}}^{1 - 2\alpha _{{\Bbb P}}}$
as $x_{{\Bbb P}}\to 0$, with fixed $x_{{\Bbb P}}s$ and $t$.
(Thus we are taking $s\to \infty $.)
This power law is roughly $x_{{\Bbb P}}^{-1}$ since
$\alpha _{{\Bbb P}}\approx 1$.  However, if one makes a jet
measurement at a given $\sqrt s$ and $E_{T}$, the cross sections in
Eq.~(\ref{eq:DiffJet}) etc.,
also have $x_{{\Bbb P}}$-dependence from
the factor
$f_{b/{\Bbb P}}(x_{b}/x_{{\Bbb P}})$ of the parton density in the
pomeron.  As $x_{{\Bbb P}}$ is decreased, $x_{b}/x_{{\Bbb P}}$
increases,
and the generally strong zero in a parton density for $x\to 1$ greatly
reduces the cross section.  Thus, even for pure pomeron exchange,
{\em one should not expect to see $x_{{\Bbb P}}^{-1}$ behavior in
the actual
hard diffractive scattering cross section at fixed $\sqrt s$.}
The most reliable way to look
for the $x_{{\Bbb P}}^{1 - 2\alpha _{{\Bbb P}}}$ behavior is by
a scaling test: increase $s$ and decrease $x_{{\Bbb P}}$ while
keeping fixed
$x_{{\Bbb P}} \, s$, $t$ and the definition of the hard scattering
({\it e.g.}, the value of $E_{T}$).
When one uses a rapidity gap signature,
the condition of fixing $t$ is replaced by an integral over $t$.

In the absence of a scaling test, one must analyze the data at
each value of $x_{{\Bbb P}}$, using
Eq.~(\ref{eq:DiffJet})
to obtain parton densities.  The parton densities should then be
consistent between different values of $x_{{\Bbb P}}$.
If not, there is some contamination by exchange of Reggeons other
than the pomeron.
Alternatively, one could analyze
the $x_{{\Bbb P}}$ dependence
of the cross section after factoring out an assumed or fitted
factor for the cross section for pomeron-induced hard
scattering.  That factor could be, for example, everything to the
right of $x_{{\Bbb P}}^{1-2\alpha _{{\Bbb P}}}$ in
Eq.~(\ref{eq:DiffJet}).

To see the pomeron clearly, one should do this analysis over a
range of $x_{{\Bbb P}}$ that includes the transition between
Reggeon and pomeron regions.  An appropriate range might be
$0.85 < 1 - x_{{\Bbb P}} < 0.98 \,$.  The range should include a
definitely non-pomeron region, as well as the region nearest to
$x_{{\Bbb P}} = 0$ that one expects to be pomeron-dominated.

The scaling test above (``Regge scaling'') is easier to do for
diffractive deep inelastic scattering.  The diffractive structure
functions are given by Eqs.~(\ref{eq:DiffF1}) and (\ref{eq:DiffF2}),
and the Regge scaling
test is to decrease $x_{\it bj}$ and $x_{{\Bbb P}}$ while holding $Q^{2}$ and
$x_{\it bj}/x_{{\Bbb P}}$ fixed.  This test can be done at fixed lepton-hadron
center-of-mass energy, even though the ideal test is to vary $s$
as well, since that avoids all questions about the ratio $F_{1}/F_{2}$.
(Then one can keep $y$ fixed,
as well as $Q^{2}$ and $x_{\it bj}/x_{{\Bbb P}}$.)

One should also analyze the $t$ dependence.  For that purpose,
parton densities in the pomeron are to be regarded as purely
non-perturbative unknown quantities.  As $|t|$
increases, it is natural that
the pomeron should become a smaller object, in analogy with the
way a virtual photon becomes more point-like as $Q$ increases.
Therefore the
parton densities could change very noticeably with $t$.
Depending on
the design of the Roman Pots, the measurements may go out to
relatively large $|t|$ by the standards of soft hadron-physics.
If so, the $t$ dependence of $\alpha _{{\Bbb P}}$ --- see
Eq.~(\ref{eq:alphaPom})
--- will have to be measured when assessing the $x_{{\Bbb P}}$-dependence.

Notice that the UA8 \cite{UA8} experiment was sensitive to $-t$
between $1$ and $2 \,{\rm GeV}^{2}$, a range that is beyond the
conventional range of fits to diffractive physics.  It is
therefore important to measure the pomeron-hadron coupling
$\beta _{p{\Bbb P}}(t)$ at these values of
$t$.  That is, {\em a program of new measurements of conventional
diffractive and elastic physics is essential for the full
implementation of our program}.
With the large $\sqrt s$ available at the colliders, the
kinematic region can go well beyond that available when
diffractive physics was a common subject of experimental
investigation.
In particular, one can go to substantially larger $-t$,
where perturbative methods for the pomeron may start to be
applicable.
This would be in the range up to a few GeV$^{2}$.
Along with
measurements of single diffractive excitation, the program
should include measurements of elastic scattering,
where there has been interesting recent theoretical work, for
example, by Sotiropoulos and Sterman \cite{SotSt}.

As to the rapidity-gap method, the Regge predictions for making
rapidity gaps in soft scattering need to be tested, since the
analysis of the hard scattering depends on the applicability of
standard Regge theory to low transverse momentum phenomena.

With elastic diffraction, where the diffracted proton is detected
(in a Roman Pot detector), a further test of the Regge behavior
can be made by also measuring the cross section when the
diffracted proton is replaced by a hadron or hadrons in a
different charge state.  This only works with a suitable
detector, of course.  The Regge exchange that gives the
diffraction can no longer have vacuum quantum numbers, so that it
cannot be a pomeron.  In formulae like Eqs~.(\ref{eq:DiffJet})
and (\ref{eq:elastic}), $\alpha _{{\Bbb P}}\approx 1$ is replaced by a smaller
value, for non-leading Reggeon exchange.

\subsection{Bjorken Scaling}
\label{subsec:check.twist}

To verify that hard diffractive scattering is as we have
described it, one must also verify that it obeys approximate
Bjorken scaling.  (Of course, QCD predicts logarithmic violations
of Bjorken scaling for hard diffractive processes, just as for
inclusive hard scattering.  But the dominant issue here concerns
the power law.)

In the case of deep inelastic scattering, a test of Bjorken
scaling involves varying $Q^{2}$, while
keeping $x_{\it bj}/x_{{\Bbb P}}$, $x_{{\Bbb P}}$ and
$t$ fixed.  The diffractive structure functions in
Eqs.~(\ref{eq:DiffF1}) and (\ref{eq:DiffF2}) should be
approximately constant.
Fixing $x_{\it bj}/x_{{\Bbb P}}$ while varying $Q^{2}$ is
exactly the test of
Bjorken scaling for deep inelastic scattering with
the pomeron treated as the target.  Fixing $x_{{\Bbb P}}$ and $t$ ensures
that there are no confounding effects from the variation of the
pomeron factors in Eqs.~(\ref{eq:DiffF1}) and (\ref{eq:DiffF2}).
When one uses a rapidity gap signature, the condition of fixing
$t$ is again replaced by an integral over $t$.

For the other processes, one makes the obvious generalizations.
For example, consider diffractive jet production in hadron-hadron
collisions.  Bjorken scaling involves increasing the jet $E_{T}$
while holding fixed the hard scattering scaling variables $x_{a}$
and $x_{b/{\Bbb P}}$, and the pomeron
variables $x_{{\Bbb P}}$ and $t$.  (Again, the
condition of fixing $t$ may be replaced by integrating over it.)
The variables $x_{a}$ and $x_{b}/x_{{\Bbb P}}$ are measured from two-jet
production, by assuming parton-model kinematics in the
pomeron-proton collision.
The cross section in Eq.~(\ref{eq:DiffJet}) will be proportional
to $1/E_{T}^{2}$.  A cross section differential in $E_{T}$,
$d\sigma /dt\, dx_{{\Bbb P}}\, dE_{T}$, would scale
as $1/E_{T}^{3}$.  There will be the
usual logarithmic violations of this scaling.
This test necessarily involves varying the beam energy, or
by bringing in more theory and using a fit of the
diffractive parton densities.

\subsection{Combination}

One test that can be done at fixed $s$ is to increase $E_{T}^{2}$ ,
while holding $x_{a}$ and $x_{b}/x_{{\Bbb P}}$ fixed.  This scaling limit has
$x_{{\Bbb P}} \propto  E_{T}^{2}$.  The cross-section
$d\sigma /dt\,dx_{{\Bbb P}}$, integrated over a range of
$E_{T}$, will scale like $1/E_{T}^{4\alpha _{{\Bbb P}}}$
from a combination of Regge and Bjorken scaling.

For DIS, the corresponding test has $Q^{2}$ being increased with
$x_{\it bj}/x_{{\Bbb P}}$ fixed, and with $x_{{\Bbb P}}$,
or equivalently $x_{\it bj}$, being
increased in proportion to $Q^{2}$.  The diffractive structure
functions in Eqs.~(\ref{eq:DiffF1}) and (\ref{eq:DiffF2}) should
scale like
\begin{eqnarray}
\frac {dF_{1}^{\rm diffractive}(x_{\it bj},Q;t,x_{\Bbb P})}{dt\,dx_{\Bbb P}}
    &\propto & \frac {1}{Q^{4\alpha _{{\Bbb P}}}} ,
\\
\frac {dF_{2}^{\rm diffractive}(x_{\it bj},Q;t,x_{\Bbb P})}{dt\,dx_{\Bbb P}}
    &\propto & \frac {1}{Q^{4\alpha _{{\Bbb P}}-2}} ,
\end{eqnarray}
in this limit with $x_{{\Bbb P}}\propto Q^{2}/s$.
The different powers of $Q$ for $F_{1}$ and $F_{2}$ reflect the
different powers of $x_{{\Bbb P}}$ in Eqs.~(\ref{eq:DiffF1}) and
(\ref{eq:DiffF2}).

%=================================================

\section {Summary of Processes to be Measured}
\label{sec:Processes}

\subsection{Hard Processes}

Here follows a list of the hard diffractive
processes that need to be investigated:
\begin{enumerate}

\item $ep \to  eXp$: diffractive DIS, as a function of $x_{\it bj}$, $Q$,
   $x_{{\Bbb P}}$, and $t$.

\item The same with 2 or more jets of transverse momentum of
   order $Q$.

\item $\gamma p\to X+{\rm jets}+p$: diffractive photoproduction of jets,
   as function of $p_{T {\rm jet}}$, $x_{{\Bbb P}}$, and $t$;
   both direct and resolved processes.

\item $p \bar p \to  p + {\rm jets} + X$:  diffractive
   hadroproduction of jets.

\item The same for $W$, $Z$ and high-$p_{T}$ production of prompt
   photons.

\item $\gamma +p \to  \rho  + {\rm jets} + X$: diffractive photoproduction of
   jets, with the photon being diffracted to a $\rho $-meson.

\item The same with the $\rho $ replaced by any other vector meson,
   {\it e.g.}, $\omega $, $\phi $, $J/\psi $.

\item All of the above with the diffracted hadron replaced by a
   system separated from the rest of the event by a rapidity gap.

\item Any of the above with the diffracted hadron replaced by
   a hadron in a different charge state, so that the process
   cannot occur by pomeron exchange.  Such processes should be
   suppressed by a power of $1/x_{{\Bbb P}}$ compared to processes where
   pomeron exchange is allowed.

\end{enumerate}

\subsection{Soft Processes}

Here follows a list of soft diffractive processes that need to be
further investigated in order to provide better quantitative
information for the soft part of hard diffractive processes:
\begin{enumerate}

\item Elastic hadron-hadron scattering is already well-known as a
   function of $s$ and $t$.

\item $p \bar p \to  p \bar p^{*}$:  Single diffraction, as a function of
   $s$, $t$, and $x_{{\Bbb P}}$.  Compare with the triple-Regge formalism.

\item $p \bar p \to  p^{*} \bar p^{*}$:  Double diffraction, with $p^{*}$,
   $\bar p^{*}$ being systems of relatively low mass which will be
   separated by a rapidity gap.  Compare with the triple-Regge
   formalism.

\item All of the above, with the $\bar p$ replaced by a photon
   $\gamma $, and the diffracted $\bar p$ replaced by a vector meson
   ($\rho $, $\omega $, $\phi $, $J/\psi $).
   One can also do this for a large virtuality
   photon and for large $t$ --- see Ref.\ \onlinecite{Brodskyetal}.

\item Any of the above with the diffracted hadron replaced by
   a hadron in a different charge state, so that the process
   cannot occur by pomeron exchange.  Such processes should be
   suppressed by a power of $1/x_{{\Bbb P}}$ compared to processes where
   pomeron exchange is allowed.

\end{enumerate}

%=================================================

\section {Conclusions}
\label{sec:concl}

We have proposed a systematic investigation of hard diffractive
scattering, with the unifying element being a
measurement of the parton densities in the pomeron, as defined by
the Ingelman-Schlein model. Since there is a prediction that this
model should fail, the results of the measurements must have
non-trivial implications for our understanding of
non-perturbative QCD.  (Even though the Ingelman-Schlein model may
fail, the UA8 data show order-of-magnitude agreement with it when
a reasonable ansatz is used for the parton densities.  Thus the
model is close enough to remain a useful basis for
planning experiments.)

By coordinating measurements between HERA and Fermilab, one can
readily measure the flavor-separated quark and gluon densities,
and then perform non-trivial tests of QCD factorization.  Since
the distributions of $u$, $d$, $\bar u$ and $\bar d$ quarks in
the pomeron are equal, two processes suffice to make the
measurement --- say DIS and direct
photoproduction of jets.  Then other
hard-scattering processes, {\it e.g.} hadroproduction of jets or
of $W$ or $Z$ bosons, with a diffractive requirement imposed,
provide tests of the picture.

In view of the possibility that factorization might hold for
deep inelastic processes, while failing for other processes,
one should perhaps attempt to treat
diffractive DIS without and with the measurement of final-state
jets as the basic processes to measure the parton densities in
the pomeron.  Since the diffractive requirement will reduce the
effective center-of-mass energy, the transverse momenta of the jets
may be so low as to make jet physics marginal.  But it is worth
trying this.  There should be enough consistency requirements
({\it e.g.}, in the $Q$ dependence) to give a test of
factorization within the DIS processes alone.

The diffractive requirement can be imposed by Roman Pot
(``quasi-elastic'', ``exclusive diffractive'' condition) or by
a simple rapidity gap requirement.  The similar rapidity coverage
of the four experiments CDF, D0, H1 and ZEUS implies that rapidity
gap measurements should be rather directly comparable between the
different experiments.  The data from UA8 on diffractive jet
production and from the other experiments with a rapidity gap
condition indicates that there will be plenty of data, with
rates on the order of $1\%$ or higher of the event rate without
the diffractive condition on the final state.

The question of whether the pomeron is dominated by gluons or
by quarks is an important issue that needs to be resolved by
experiment, as is the issue of whether or not the momentum sum rule
is valid.  The measurements we propose will answer these
questions.

The $t$-dependence of the parton densities in the pomeron is also of
considerable interest.  It can only be probed using the Roman Pot
technique.

CFS \cite{CFS} predict that the
Ingelman-Schlein picture will break down, in that there will be,
in certain processes,
an excess of events at $x_{b}/x_{{\Bbb P}}\approx 1$.
These should exhibit Bjorken scaling
when the characteristic momentum of the hard scattering is
increased.  The excess should be present in
hadroproduction of jets, and appears to have been observed by
UA8 \cite{UA8}.  In a process such as deep inelastic scattering
or in the ``direct'' (as opposed to ``resolved'') component of
jet photoproduction, the excess, if any, is
predicted to be
of higher twist, {\it i.e.}, suppressed by a power of
$1/Q$ ($1/p_{T}$, in the case of photoproduction).  The systematics of
a breakdown of
factorization in diffractive processes, as a function of the
process and of the kinematic variables, in particular $t$,
will provide important information on the space-time structure of
the pomeron.

Since part of the formulae for the cross sections involve
conventional Regge theory, it is important to test Regge theory
further
at the energies now available, particularly with regard to the
rapidity gap cross sections \cite{O,UA4}.

Another possible test of the Regge factorization could be made using two-jet
production in double pomeron exchange (DPE) interactions at the Tevatron.
This could be done using the rapidity-gap method.  In the case of D0,
signals in their
forward and backward scintillation counters have been required as
a part of all triggers, thereby preventing triggering on DPE event.  Those
counters, and similar ones in CDF,
could in fact be used in veto mode to trigger on
DPE events \cite{DPE}.
The CFS argument should also apply to jet production in DPE.

%====================================================
\section*{Acknowledgments}

This work was supported in part by the U.S. Department of Energy
under grant DE-FG02-90ER-40577, by the U.S. National Science
Foundation, and by the Texas National
Laboratory Research Commission grant to the CTEQ collaboration.
One of the authors (JJW) wishes to acknowledge partial support
from the DESY Directorate in 1993.94.
We would
like to thank
many colleagues for discussions, notably our colleagues
in the
CTEQ collaboration, and L. Frankfurt, E.M.
Levin, and M. Strikman.  We also wish to thank the anonymous
referee for a number of thoughtful comments that have improved
the content and clarity of the paper.

%==========================================

\appendix

\section{Light-Cone Coordinates}
\label{sec:lc}

In this appendix, we define light-cone
coordinates.  Then we show how they can be used to make
measurements of, for example, the momentum of an exchanged
pomeron from the hadronic final-state without observation of the
diffracted proton.

\begin{figure}
   \begin{center}
      \leavevmode
      \epsfxsize=0.4\hsize
      \epsfbox{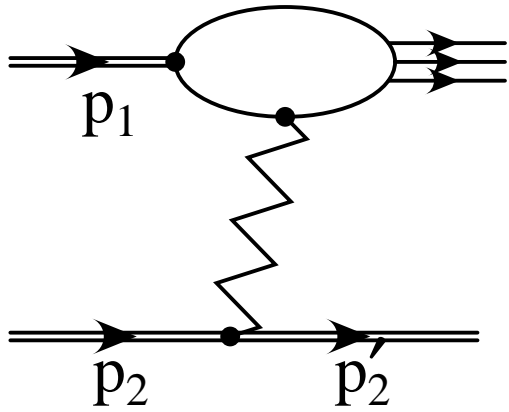}
   \end{center}
   \caption{Reaction used in definition of light-cone
      coordinates.
   }
   \label{fig:reaction}
\end{figure}

Consider the reaction shown in Fig.\ \ref{fig:reaction} where
incoming particle $p_{2}$
(traveling in the $+z$ direction) emits a pomeron ${\Bbb P} $ which then
interacts with particle $p_{1}$ (traveling in the $-z$ direction).
The diffracted particle $p_{2}'$ emerges containing a high
fraction $x$
($\gtrsim 0.95$) of the momentum of
particle $p_{2}$.  The pomeron carries a
momentum fraction $x_{{\Bbb P}} = 1-x$.  The direction of the $+z$
axis is the same as in the ZEUS detector when particle $p_{2}$ is
the proton.

One way of determining the momentum fraction of the pomeron
({\it\`a la} UA8)
is to measure the momentum of the diffracted particle $p'_{2}$, using
Roman Pot detectors.  Currently, none of the collider detectors at
the Tevatron are equipped with Roman Pot detectors.  But this
should be considered as an upgrade option.  The ZEUS experiment
recorded some data in 1993 with some partially equipped Roman Pot
detectors.

It is also of interest to measure the cross section when the
diffracted particle is replaced by a system of relatively low
mass.

Another determination of $x_{{\Bbb P}}$ can be carried out by
measuring the momenta of the other particles $q_{1}$,\dots,$q_{n}$ in
the final-state.  When these are expressed in light-cone
coordinates, it is easy to show that there is a rapidity gap
on the side of the detector towards which $p_{2}$ travels (the $+z$
direction).  The computation of $x_{{\Bbb P}}$ is dominated by the
particles close to the edge of the rapidity gap and by particles
of high transverse momentum, as we will also show.
The same
method is already used to reconstruct the photon kinematics
from the hadron final-state in electron-proton scattering
\cite{RapGapZEUS,JacquetBlondel}.

\begin{figure}
   \begin{center}
      \leavevmode
      \epsfxsize=0.8\hsize
      \epsfbox{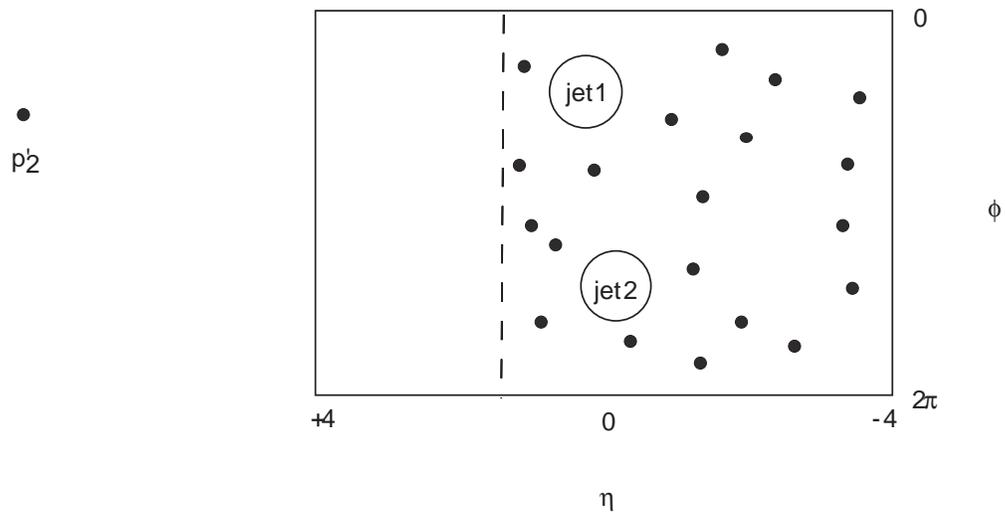}
   \end{center}
   \caption{$\eta $-$\phi $ plot for the hard diffractive production of 2
       jets.
   }
   \label{fig:etaphi}
\end{figure}

The energy and $z$ component of the momentum of each of the
outgoing particles $q_{i}$ can be written
\begin{eqnarray}
   q^{0}_{i} &=& m^{T}_{i} \cosh y_{i} ,
\nonumber\\
   q^{z}_{i} &=& m^{T}_{i} \sinh y_{i} ,
\end{eqnarray}
where
$m^{T}_{i} \left(= \sqrt {{q^{T}_{i}}^{2} + m_{i}^{2}}\right)$
is the transverse
mass, and $y_{i}$ is the rapidity of the particle.
We can define the following coordinates
\begin{eqnarray}
   {q_{i}}^{+} &=& \frac {q^{0}_{i} + q^{z}_{i}}{\sqrt 2}
          = m^{T}_{i} e^{yi} / \sqrt 2 ,
\nonumber\\
   {q_{i}}^{-} &=& \frac {q^{0}_{i} - q^{z}_{i}}{\sqrt 2}
          = m^{T}_{i} e^{-y_{i}} / \sqrt 2 ,
\nonumber\\
  {q_{i}}^{T} &=& ({q_{i}}^{x},{q_{i}}^{y}) ,
\label{eq:lfdefd}
\end{eqnarray}
which are called light-cone coordinates.
When the mass of
a particle is small compared with its transverse momentum, the
rapidity is well approximated by the pseudorapidity $\eta _{i}$, which
is determined directly from the polar angle $\theta $:
$\eta = - \ln \tan (\theta /2)$.
So it is useful to
consider the $\eta $-$\phi $ plot, which is shown in
Fig.~\ref{fig:etaphi}, for the hard diffractive production of 2
jets.

As above, $x$ is the fraction of the momentum
taken by the outgoing particle $p'_{2}$.  This is most conveniently
defined in term of light-cone coordinates:
\begin{equation}
   x = \frac {{p_{2}'}^{+}}{{p_{2}}^{+}} .
\end{equation}
Then momentum conservation, $p_{2} = p'_{2} + \sum q_{i} - p_{1}$, yields
\begin{eqnarray}
    x_{{\Bbb P}} &\equiv & \frac {{p_{2}}^{+} - {p'_{2}}^{+}}{{p_{2}}^{+}}
\nonumber\\
   &=& \frac {\sum  {q_{i}}^{+} - {p_{1}}^{+}}{{p_{2}}^{+}} .
\end{eqnarray}
Since ${p_{1}}^{+}/{p_{2}}^{+} = {\cal O}(m^{2}/s)$,
we can neglect ${p_{1}}^{+}$,
except if the
event is almost exactly elastic, which is not the case when there
is a hard scattering.
Hence
\begin{equation}
   x_{{\Bbb P}} \approx
   \sum  \frac {m^{T}_{i} e^{y_{i}}}{E_{2} + {p_{2}}^{z}}
   \approx
   \sum _{i}
   \frac {\sqrt {{q^{T}_{i}}^{2} + m_{i}^{2}}e^{-(y_{p_{2}}-y_{i})}}{m_{p}} .
\label{eq:x.pom.from.q}
\end{equation}
For diffraction, we require $x_{{\Bbb P}}$ to be small.
Then Eq.~(\ref{eq:x.pom.from.q}) shows that the rapidities of all
the particles $q_{i}$ must be substantially less than the
rapidity of the incoming proton $p_{2}$.  Hence we must have a
rapidity gap, and to the extent that there are few particles of
very low transverse momentum, this implies a gap in
pseudorapidity, as we claimed earlier.
It is important to have calorimetry coverage over as large a
rapidity interval as possible in order to confirm the presence of
this gap.

Moreover, only those particles with the largest $q^{T}$ and/or
largest rapidity contribute significantly to the sum.
Substituting pseudorapidity ($\eta _{i}$) for rapidity ($y_{i}$) should
result in a small error in the determination of $x_{{\Bbb P}}$.
Hence, Eq.~(\ref{eq:x.pom.from.q}) gives an effective method for
the measurement of $x_{{\Bbb P}}$.  Note that the lack of coverage in the
detector on the side {\it opposite} to the diffracted hadron is
irrelevant.

In a typical ``minimum-bias'' final state, there is an
approximately uniform distribution of hadrons in $\eta $  and $\phi $ in
the non-gap region.  Hence
$x_{{\Bbb P}} = {\cal O}(e^{-\eta _{{\rm gap}}})$,
where $\eta _{{\rm gap}}$ is the difference
between the (pseudo-)rapidity of the edge of the gap and the
rapidity of the proton.
For example, if $\eta _{{\rm gap}} = -3$,
$x_{{\Bbb P}}$ is less than or equal to .05.

%==========================================

\section{Diffractive Structure Functions}
\label{sec:DiffStr}

The right-hand sides of (\ref{eq:DiffF1}) and (\ref{eq:DiffF2})
are, of course, written on the assumption that the
Ingelman-Schlein model is valid.  Within the model, structure
functions in the pomeron are defined in exactly the same way as
structure functions in the proton.

One of our intentions is to test the Ingelman-Schlein model.  So
the diffractive structure functions $F_{i}^{{\rm diffractive}}$ on the
left-hand sides of these equations need to be defined.
That is
the purpose of this appendix.

A diffractive DIS process is just semi-inclusive DIS, with the
final state restricted to a
particular kinematic region for the detected outgoing hadron.  So
there are actually five structure functions\cite{Ravndal}, only
four of which contribute to the unpolarized cross section.
What we will show is that the cross section is almost certainly
dominated by two of these, which in Kingsley's\cite{Kingsley}
conventions are like the usual $F_{1}$ and $F_{2}$.  Kingsley's
conventions are more convenient for our purposes than those of
Meng et al.\cite{Meng}.  We write
\begin{eqnarray}
    W^{\mu \nu } &=& \frac {1}{4\pi } \sum _{X}\int
            \langle P| J^{\nu }(0) |P', X\rangle
            \langle p', X| ^{J}(0) |P\rangle
\nonumber\\
 &=& \left(-g^{\mu \nu }+\frac {q^{\mu }q^{\nu }}{q^{2}} \right) V_{1}
   + \frac {P^{\mu }P^{\nu }}{p\cdot q} V_{2}
\nonumber\\
  && + \frac {{P'}_{T}^{\mu }{P'}_{T}^{\nu }}{p\cdot q} V_{3}
   + \frac {{P'}_{T}^{\mu }P^{\nu } + P^{\mu }{P'}_{T}^{\nu }}{2 p\cdot q}
     V_{4}
   + i \frac {{P'}_{T}^{\mu }P^{\nu } - P^{\mu }{P'}_{T}^{\nu }}{2 p\cdot q}
     V_{5},
\label{eq:DiffW}
\end{eqnarray}
where $P^{\mu }=p^{\mu }-p\cdot q/q^{2}$, and ${P'}_{T}^{\mu }$ is
component  of the momentum of the
outgoing diffracted hadron that is transverse to both $p$ and
$q$.  In the diffractive limit, ${P'}_{T}^{\mu }$ is the
component of the momentum of the outgoing hadron that is
transverse to the collision axis.
We have written the scalar structure functions in terms of
dimensionless structure functions.  The first two structure
functions have exactly the same form as for inclusive deep
inelastic scattering.  We have used symbols $V_{i}$ to avoid
confusion with the usual structure functions $F_{i}$.  The fifth
structure gives a zero contribution when contracted with the
symmetric leptonic tensor in unpolarized scattering.

Normal power counting for the hard scattering shows that each of
the $V_{i}$ scales in the Bjorken limit.  This power counting
applies before the more detailed arguments needed to get
factorization are applied; it is really dimensional counting of
powers of the large mass scale in the problem.  Since there is a
factor of a small transverse momentum with the $V_{3}$ and $V_{4}$
structure functions, they give a nonleading power when contracted
with the leptonic tensor.  So to a good approximation, only $V_{1}$
and $V_{2}$ contribute.

When we contract with the leptonic tensor, the dependence on the
variables $x_{\it bj}$ and $y\equiv q\cdot p/l\cdot p$
is the same as in inclusive
deep inelastic scattering.  The cross section also depends on the
variables $t$ and $x_{{\Bbb P}}$ of the diffracted hadron.  By
multiplying $V_{1}$ and $V_{2}$ by a factor we may therefore write the
cross section as
\begin{equation}
   \frac {d\sigma }{dx \, dy \, dt \, d_{{\Bbb P}}}
   = \frac {4\pi \alpha ^{2}s}{Q^{4}}
     \left[ xy^{2} \frac {dF^{{\rm diffractive}}_{1}}{dt \,  d_{{\Bbb P}}}
            (1-y) \frac {dF^{{\rm diffractive}}_{2}}{dt \,  d_{{\Bbb P}}}
     \right].
\label{eq:Diffsigma}
\end{equation}
This defines the diffractive structure functions in Eqs.\
(\ref{eq:DiffF1}) and (\ref{eq:DiffF2}) in terms of the $y$
dependence of the cross section.  Note that although we use the
label ``diffractive'' on these structure functions, we do not
mean to imply that they are due to diffraction (i.e., pomeron
exchange).  Rather the label only indicates that they are for a
process and kinematic configuration that is appropriate for
investigating diffraction.

%==========================================

\section{Normalization}
\label{sec:norm}

\begin{figure}
   \begin{center}
      \leavevmode
      \epsfbox{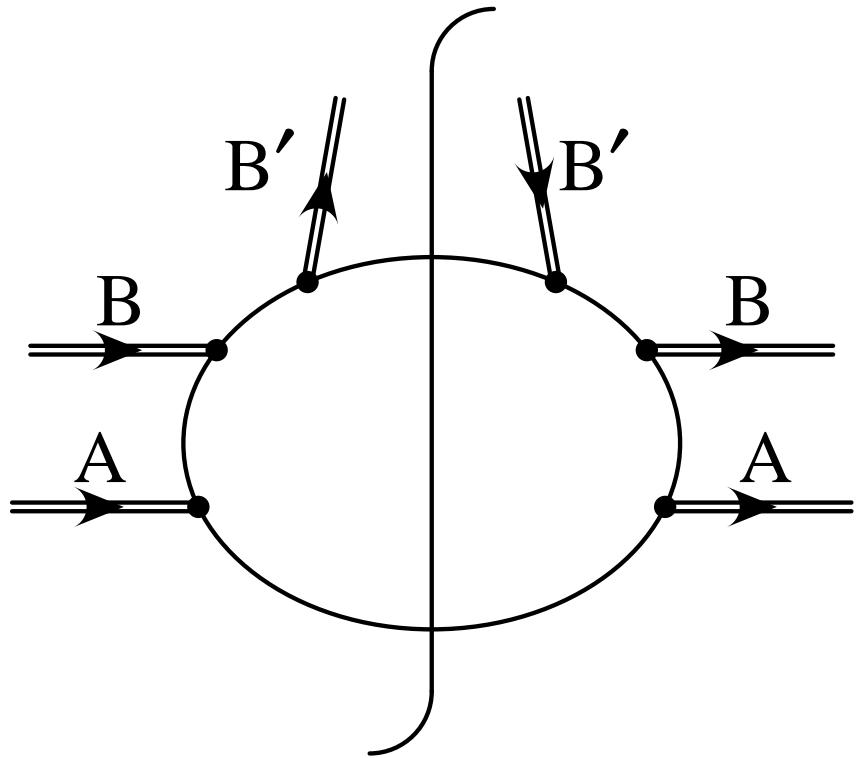}
   \end{center}
   \caption{Diffractive scattering is given by this discontinuity
      of a six point function.
   }
   \label{fig:Disc}
\end{figure}

In this section we provide some arguments about the normalization
of Eq.~(\ref{eq:DiffJet}).  We are considering inclusive cross
sections for the process $A+B \to  X+B'$.  Elementary manipulations
show that the cross section differential in the variables for
$B'$ is
\begin{equation}
   \frac {d\sigma (A+B \to  X + B')}{dt \, dx_{{\Bbb P}}} =
   \frac {1}{32\pi ^{2}s}
   \frac {1}{i} {\rm disc} {\cal M} ,
\label{eq:SD}
\end{equation}
where ${\rm disc} {\cal M}$ is the appropriate discontinuity of
the amplitude shown in Fig.~\ref{fig:Disc}. We have made
approximations valid for $x_{{\Bbb P}} \ll 1$ and $s \gg m^{2}$.

To get a jet cross
section, as in Eq.~(\ref{eq:DiffJet}), one would restrict the
integral over the final state $X$.
But suppose first that one integrates over all $X$, to get the
normal single diffractive cross section.  Moreover, let us take
the triple Regge limit: $m^{2} \ll M_{X}^{2} \ll s$ with $t$ fixed.  (Note
that $x_{{\Bbb P}} = M_{X}^{2}/s$.)  Then the triple Regge formula for the
cross section is
\begin{equation}
   \frac {d\sigma (A+B \to  X + B')}{dt \, dx_{{\Bbb P}}} =
   \frac {1}{16\pi } |\beta _{B{\Bbb P}}(t)|^{2} \,
       |\xi (\alpha _{{\Bbb P}}(t))|^{2} \,
       x_{{\Bbb P}}^{1-2\alpha _{{\Bbb P}}(t)} \,
       G_{{\Bbb PPP}}(t) \,
       \beta _{A{\Bbb P}}(0) \,
       M_{X}^{2\alpha _{{\Bbb P}}(0)-2} ,
\label{eq:Pomeron3}
\end{equation}
where we have used the normalizations of Ref.~\cite{FieldFox}.
The signature factor
\begin{equation}
   \xi (\alpha _{{\Bbb P}}) =
   - \frac {\tau +e^{-i\pi \alpha _{{\Bbb P}}}}{\sin \pi \alpha _{{\Bbb P}}},
\label{eq:Signature}
\end{equation}
is close to $i$ for the pomeron, which has even signature $\tau =1$.

To get a Pomeron-proton cross-section, one naturally would factor
out the factors associated with the $B{\Bbb P}B'$ vertex.
According to \cite{FieldFox} the result is
\begin{equation}
   \sigma _{{\rm tot}}(A{\Bbb P}) =
       G_{{\Bbb PPP}}(t) \,
       \beta _{A{\Bbb P}}(0) \,
       M_{X}^{2\alpha _{{\Bbb P}}(0)-2} ,
\label{eq:pomprot}
\end{equation}
which is exactly the formula one would use for the total
cross section for hadron-hadron scattering, with the
triple-pomeron coupling $G_{{\Bbb PPP}}(t)$
replacing a pomeron-hadron coupling.  The formulae given by
Kaidalov\cite{Kaidalov} are completely consistent with the above.

Now, in both \cite{FieldFox} and \cite{Kaidalov},
Eq.~(\ref{eq:Pomeron3}) and Eq.~(\ref{eq:pomprot}) are claimed to
be valid when the pomeron exchanged with the $B$ particle is
replaced by any other Regge trajectory, as would be appropriate
for a charge exchange process, for example.

One could argue that the physical region for
diffraction is a long way from any particle pole on the pomeron
trajectory, and thus that it is quite ambiguous as to how to
factor off the pomerons to go from Eq.~(\ref{eq:Pomeron3}) to
Eq.~(\ref{eq:pomprot}).  But suppose one has a theory with a
massless scalar particle $R$.  Then one can construct the same
diffractive process as in Eq.~(\ref{eq:Pomeron3}), but with quantum
numbers for $B$ and $B'$ such that the leading term in the cross
section has $R$ exchange instead of pomeron exchange.  Then
Eq.~(\ref{eq:Pomeron3}) is replaced by
\begin{equation}
   \frac {d\sigma (A+B \to  X + B')}{dt \, dx_{{\Bbb P}}} =
   \frac {1}{16\pi } |\beta _{BRB'}(t)|^{2} \,
       |\xi (\alpha _{R}(t))|^{2} \,
       x_{{\Bbb P}}^{1-2\alpha _{R}(t)} \,
       G_{RR{\Bbb P}}(t) \,
       \beta _{A{\Bbb P}}(0) \,
       M_{X}^{2\alpha _{{\Bbb P}}(0)-2} .
\label{eq:Regge3}
\end{equation}

The Regge pole for $R$ will be of even signature, and close to
$t=0$ we can write
\begin{equation}
   \alpha _{R}(t) = \alpha '_{R} t .
\label{eq:TrajR}
\end{equation}
The signature factor will therefore have a pole at $t=0$:
\begin{equation}
   \xi (\alpha _{R}(t)) \to  \frac {-2}{\pi \alpha '_{R} t}  \hbox{as }t\to 0.
\label{eq:SignPole}
\end{equation}
This pole is at the edge of the physical region for diffraction,
so we may use the LSZ reduction method to obtain the cross
section for $AR$ scattering.  First we obtain the discontinuity
in Eq.~(\ref{eq:SD}) from Eq.~(\ref{eq:Regge3}).  Then for each
of the two $R$ exchanges, we have to remove the factor $1/t$
and the square root of the residue in $BB'$ elastic scattering:
\begin{equation}
   {\cal M}_{BB'\to BB'}
   = \beta _{BRB'}(t)^{2} \xi (\alpha _{R}(t)).
\label{eq:BBprime}
\end{equation}

The result is that the discontinuity at $t=0$ of the $AR$ elastic
scattering amplitude is
\begin{equation}
  {\rm disc}{\cal M}(t=0, AR\to AR)
  = \frac {4i}{\alpha '_{R}} s^{\alpha _{{\Bbb P}}(0)}
    G_{RR{\Bbb P}}(0) \beta _{A{\Bbb P}}(0) ,
\label{eq:ARdisc}
\end{equation}
where $M_{X}^{2}$ in Eq.~(\ref{eq:Regge3}) is now replaced by $s$.
The corresponding total cross section is
\begin{equation}
  \sigma _{{\rm tot}}(AR)
  = \frac {2}{\alpha '_{R}} s^{\alpha _{{\Bbb P}}(0)-1}
    G_{RR{\Bbb P}}(0) \beta _{A{\Bbb P}}(0) .
\label{eq:ARsigma}
\end{equation}
This is in contradiction with Eq.~(\ref{eq:pomprot}), which is
supposed to be valid generally, and not just for the
pomeron-proton cross section.

%==================================================

%======= \listoffigures

%==================================================

%===\clearpage

\end{document}